\documentclass[10pt]{iopart}
\usepackage{iopams,epsfig}  
\begin{document}

\title{Review of the results of the KaoS Collaboration}

\author{A. F{\"o}rster for the KaoS-Collaboration:\\
        I. B{\"o}ttcher$^{d}$, F. Dohrmann$^{f}$, 
        A. F{\"o}rster$^{a,*}$, E. Grosse$^{f,g}$,\\ 
        P. Koczo{\'n}$^{b}$, 
        B. Kohlmeyer$^{d}$, S. Lang$^{a}$, F. Laue$^{b,\dag}$, \\
        M. Menzel$^{d}$, 
        L. Naumann$^{f}$, H. Oeschler$^{a}$, 
        M. P{\l}osko{\'n}$^{b}$, \\ 
        F. P{\"u}hlhofer$^{d}$, W. Scheinast$^{f}$, 
        A. Schmah$^{a}$, T. Schuck$^{c,\ddag}$, \\ 
        E. Schwab$^{b}$, 
        P. Senger$^{b}$, Y. Shin$^{c}$, H. Str{\"o}bele$^{c}$, \\ 
        C.Sturm$^{a}$, 
        F. Uhlig$^{a}$, A. Wagner$^{f}$, W. Walu{\'s}$^{e}$}

\address{$^{a}$ Technische Universit{\"a}t Darmstadt,
                D-64289 Darmstadt, Germany}
\address{$^{b}$ Gesellschaft f{\"u}r Schwerionenforschung, 
                D-64291 Darmstadt, Germany} 
\address{$^{c}$ Johann Wolfgang Goethe Universit{\"a}t,
                D-60325 Frankfurt am Main, Germany}
\address{$^{d}$ Phillips Universit{\"a}t, 
                D-35037 Marburg, Germany} 
\address{$^{e}$ Uniwersytet Jagiello{\'n}ski, 
                PL-30-059 Krak{\'o}w, Poland}
\address{$^{f}$ Forschungszentrum Rossendorf,
                D-01314 Dresden, Germany}
\address{$^{g}$ Technische Universit{\"a}t Dresden,
                D-01062 Dresden, Germany}
\address{$^{*}$ Present address: CERN, CH-1211 Geneva, Switzerland}
\address{$^{\dag}$ Present address: Brookhaven National Laboratory, USA}
\address{$^{\ddag}$ Present address: MPI f{\"u}r Kernphysik, 
                    D-69117 Heidelberg, Germany}

\ead{Andreas.Foerster@cern.ch}

\begin{abstract}
The production of K$^+$  and of K$^-$ mesons 
in heavy-ion collisions at beam energies 
of 1 to 2~AGeV has systematically been investigated 
with the Kaon Spectrometer KaoS.
The ratio of the K$^+$ production excitation function 
for \mbox{Au+Au}  and for \mbox{C+C} reactions
increases with decreasing beam energy, which is expected 
for a soft nuclear equation-of-state.
A comprehensive study of the 
K$^+$  and of the K$^-$ emission as a function of the 
size of the collision system, of the collision 
centrality, of the kaon energy, and 
of the polar emission angle has been performed.
The K$^-$/K$^+$ ratio is found to be nearly
constant as a function of the collision centrality
and can be explained by the dominance of
strangeness exchange. On the other hand the spectral
slopes and the polar emission patterns are different for K$^-$ and
for K$^+$.
Furthermore the azimuthal distribution of the particle emission
has been investigated.
K$^+$ mesons and pions are emitted preferentially
perpendicular to the reaction plane as well in
Au+Au as in Ni+Ni collisions. In contrast
for K$^-$ mesons in Ni+Ni reactions an in-plane flow was
observed for the first time at these incident enegies.
\end{abstract}





\section{Introduction} \label{introduction}

Heavy-ion collisions provide the unique possibility to study
baryonic matter well above saturation density. The conditions
inside the dense reaction zone and the in-medium properties of
hadrons can be explored by measuring the particles created in such
collisions \cite{aich,brown}. 
In particular, strange mesons are considered to be sensitive to
in-medium modifications. Theory predicts a repulsive
K$^+$N potential and an attractive K$^-$N potential in dense
matter \cite{schaffner}.
At beam energies of 1 to 2~AGeV
strange mesons  are produced below or close to their 
respective threshold in binary nucleon-nucleon collisions 
($\rm{NN} \rightarrow \rm{K^+\Lambda N}$ at $E_{beam} = 1.6$~GeV, 
$\rm{NN} \rightarrow \rm{K^+ K^- NN}$ at  $E_{beam} = 2.5$~GeV).
The production at these energies  requires 
multiple nucleon-nucleon collisions or 
secondary collisions like e.g. $\rm{\pi N} \rightarrow \rm{K^+ Y}$
($\rm{Y} = \Lambda, \Sigma$) for the K$^+$ or 
the strangeness exchange reaction 
$\rm{\pi Y} \rightarrow \rm{K^- N}$ for the K$^-$.
K$^+$ mesons contain an $\rm{\bar{s}}$-quark and hence can hardly be
absorbed in hadronic matter consisting almost entirely of u- and 
d-quarks.
The K$^-$ on the other hand can be reabsorbed 
via the inverse direction of the strangeness 
exchange reaction mentioned above.

The experiments were performed with the Kaon Spectrometer (KaoS)
at the heavy-ion synchrotron (SIS) at the GSI in Darmstadt
\cite{senger}. The magnetic spectrometer has a large acceptance in
solid angle and in momentum ($\Omega \approx 30$~msr,
$p_{max}/p_{min} \approx 2$).  The particle identification and the
trigger are based on separate measurements of the momentum and
of the time-of-flight. 
The background due to
spurious tracks and pile-up is removed by a trajectory
reconstruction based on three large-area multi-wire proportional
counters. The short distance of 5 - $6.5$~m from the target to the focal
plane minimizes the number of 
kaon-decays in flight. The loss of
kaons by decay is accounted for by Monte Carlo
simulations using the GEANT code.
Three different collision systems have been investigated
(\mbox{Au+Au}, \mbox{Ni+Ni}, and \mbox{C+C}) at 
incident beam energies ranging from $0.6$~AGeV to $2.0$~AGeV.
Results have been published in 
\cite{misko,ahner,barth,shin,laue,menzel,sturm,foerster,uhlig}
and will in part be presented in this contribution.

Section \ref{eos} briefly summarizes 
the results on the excitation function of the 
K$^+$ production and the conclusions 
that can be drawn on the nuclear equation-of-state.
In section \ref{mapart}
the dependence of the K$^+$ and of the K$^-$
production on the collision centrality is discussed.
Section \ref{dynamics} shows details
on energy spectra and polar angle distributions
and in section \ref{azimuthal} 
azimuthal distributions for $\pi$, K$^+$, and K$^-$
are presented.


\section{K$^{\boldsymbol{+}}$ production as a probe 
         for the nuclear equation-of-state}  \label{eos}

Early transport calculations predicted that the K$^+$ yield in 
\mbox{Au+Au} 
collisions at beam energies below the production threshold 
in nucleon-nucleon collisions would be enhanced by a factor of about 
2 if a soft rather than a hard equation-of-state is assumed 
\cite{aich,li_ko}. Recent calculations take into account  
modifications of the kaon properties within the dense 
nuclear medium. 
The repulsive K$^{+}$N potential assumed depends nearly (or less than)
linearly on the baryonic density \cite{schaffner} 
and thus reduces the K$^+$ yield accordingly. On the other hand, at 
subthreshold beam energies the K$^{+}$ mesons are created in secondary 
collisions involving two or more particles and hence their production  
depends at least quadratically on the density. 
To disentangle these two competing effects we have studied 
the K$^+$ production in a light (\mbox{C+C}) and in 
a heavy collision 
system (\mbox{Au+Au}) at different beam energies near threshold. 
The maximum baryonic density reached in \mbox{Au+Au}
reactions is significantly higher than in 
\mbox{C+C} reactions.
Moreover, the maximum baryonic density reached in Au+Au reactions  
depends strongly on the compression modulus of nuclear matter $\kappa$
\cite{aichelin,li_ko} whereas in C+C collisions this dependence is
rather weak \cite{fuchs}. 
Hence, the ratio of the K$^+$ multiplicity per nucleon $M/A$
in \mbox{Au+Au} to the one in \mbox{C+C} is expected 
to be sensitive to the compression modulus $\kappa$
while providing the advantage that uncertainties 
within the experimental data (beam normalization etc.) 
and within the transport model
calculations (elementary cross sections etc.) are partly cancelled.

\begin{figure}[h]
  \begin{center}
    \epsfig{file=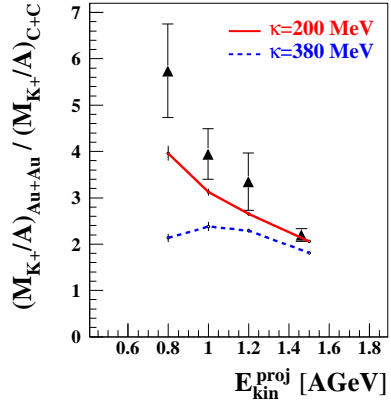,width=6.0cm}
  \end{center} 
  \caption{ The ratio of the K$^+$ multiplicitiy per nucleon 
  $M/A$ in \mbox{Au+Au} to the one in 
  \mbox{C+C} reactions as a function of the beam energy for 
  inclusive reactions. The data are compared to the results of RQMD 
  transport model calculations \cite{fuchs}. The calculations
  were performed with two different values for the compression modulus: 
  $\kappa$ = 200~MeV (a ``soft'' equation-of-state), 
  denoted by the solid line, and 
  $\kappa$ = 380~MeV (a ``hard'' EoS), indicated by the dashed line.}
  \label{fig_eos}
\end{figure}

Figure \ref{fig_eos} shows a comparison of the ratio 
$(M/A)_{\rm{Au+Au}}/(M/A)_{\rm{C+C}}$ for K$^+$ together with the predictions of 
RQMD transport model calculations \cite{fuchs}. 
The calculations were performed with two values for the compression 
modulus: $\kappa=200$~MeV (solid line) and $\kappa=380$~MeV (dashed line) 
corresponding to a ``soft'' or to a ``hard'' nuclear equation-of-state, 
respectively. The RQMD transport model takes into account  
a repulsive K$^{+}$N-potential and uses 
momentum-dependent Skyrme forces to
determine the compressional energy per nucleon (i.e. the energy stored
in the compression).
The comparison 
demonstrates clearly that only the calculation based on a soft nuclear 
equation-of-state reproduces the trend of the experimental data.
A similar result is obtained using the transport code IQMD \cite{har02}.


\section{The centrality dependence}  \label{mapart}

The centrality of the collision was derived from the multiplicity of
charged particles measured in the interval $12^{\circ} <
\theta_{lab} < 48^{\circ}$ by a hodoscope consisting of 84
plastic-scintillator modules.
In order to study the centrality dependence the data
measured close to midrapidity ($\theta_{lab} = 40^{\circ}$) 
were grouped into
five centrality bins both for \mbox{Ni+Ni} and for 
\mbox{Au+Au} collisions at $1.5$~AGeV. 

The inclusive kaon multiplicity  
for each centrality bin is defined as
$M = \sigma_{\rm{K}}/(f\cdot\sigma_R)$ with $\sigma_{\rm{K}}$ 
being the kaon production cross
section and $(f\cdot\sigma_r)$ being the fraction of the 
reaction cross-section for the
particular event class
which was determined by normalising the
charged particle multiplicity distribution measured
with a minimum bias trigger. 
The corresponding mean number of participating nucleons
for each centrality bin 
$A_{part}$ was calculated from the measured reaction
cross-section fraction for this bin using a Glauber model.

\begin{figure}
  \begin{center}
  \mbox{\epsfig{figure=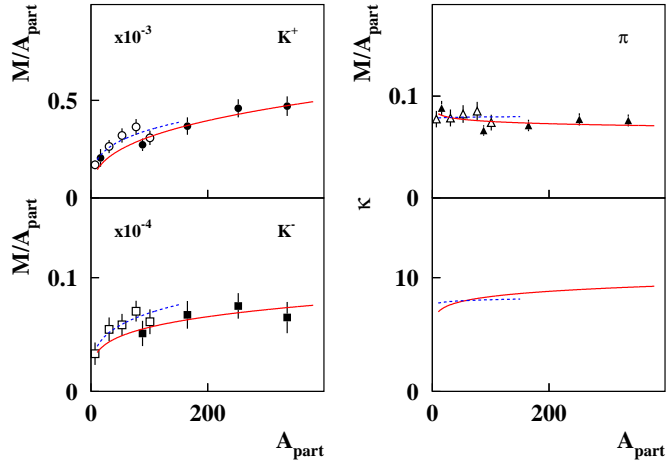,width=9cm,clip=}}
  \caption{Particle yields $M$ per participant $A_{part}$ as a
           function of $A_{part}$ obtained
           from Ni+Ni (open symbols) and
           Au+Au collisions  (full symbols) at $1.5$~GeV. 
           The lines are functions  $M \propto A_{part}^\alpha$ 
           fitted to the
           data (solid lines for \mbox{Au+Au}, 
           dashed lines for \mbox{Ni+Ni} collisions)
           with $\alpha \approx 1.2 - 1.3$. The lower right 
           panel shows the
           resulting values of the equlibration constant  
           $\kappa$ for the strangeness exchange reaction
           (see text) as determined
           from the fits shown above \cite{massaction}. 
           The data are taken around midrapidity
           and extrapolated to $4\pi$ \cite{foerster}.}
  \label{fig_ratio}
\end{center} 
\end{figure}

Figure \ref{fig_ratio} presents the 
multiplicity per number of participating nucleons 
$M/A_{part}$ as a function
of $A_{part}$ for K$^+$,
for K$^-$, and for pions.
The lines are functions  $M \propto A_{part}^\alpha$ 
fitted to the data yielding similar values of
$\alpha \approx 1.2 - 1.3$ for K$^+$ and for K$^-$
as well in \mbox{Au+Au} as in \mbox{Ni+Ni} and
$\alpha \approx 1.0$ for pions.
The resulting
K$^-$/K$^+$ ratio is about 0.02 for \mbox{Au+Au} as well as for 
\mbox{Ni+Ni} and is rather constant as a function of the 
collision centrality. 

At low incident energies the strangeness-exchange reaction
$\pi + \rm{Y} \, \rightleftharpoons  \,  \rm{K}^- + \rm{N}$
plays a key role in the K$^-$  production. 
If the rates for the K$^-$ production are equal to those for the K$^-$
absorption, this  reaction is in chemical
equilibrium. In this case the law-of-mass action is applicable
and leads to the following  relation between particle yields
$[\pi] \cdot [\rm{Y}] / [\rm{K}^-] \cdot [\rm{N}] \, = \, \kappa$
with $[\rm{x}]$ being the multiplicity of particle x and $\kappa$ the
equilibration constant \cite{massaction,oeschler}.
At this low beam energies the hyperons are
produced together with K$^+$ and K$^0$ in equal rates, hence, 
$[\rm{Y}] = [\rm{K}^+] + [\rm{K}^0] \approx 2 \cdot [\rm{K}^+]$.
The multiplicity of nucleons [N]  can be
related to the average number of participants $A_{part}$.
The pion density [$\pi$] 
contains unequal contributions from [$\pi^+$], [$\pi^0$] and
[$\pi^-$] due to the isospin  asymmetry of the colliding system.
The details of the calculation to obtain the equilibration constant
$\kappa$ from the measured data can be found in \cite{massaction}.
The results using the separate fits to the \mbox{Au+Au} and
\mbox{Ni+Ni} data are presented in the lower right panel 
of figure \ref{fig_ratio} and they show that 
$\kappa$ is independent of the centrality and of the
collision system within the experimental uncertainties.


\section{Energy spectra and polar angle distributions}  \label{dynamics}

Although the production of K$^+$  and of
K$^-$ mesons is strongly coupled via the strangeness exchange
reaction significant differences 
between K$^+$ and K$^-$ have been found \cite{foerster}.

\begin{figure}
  \begin{center}
    \mbox{\epsfig{figure=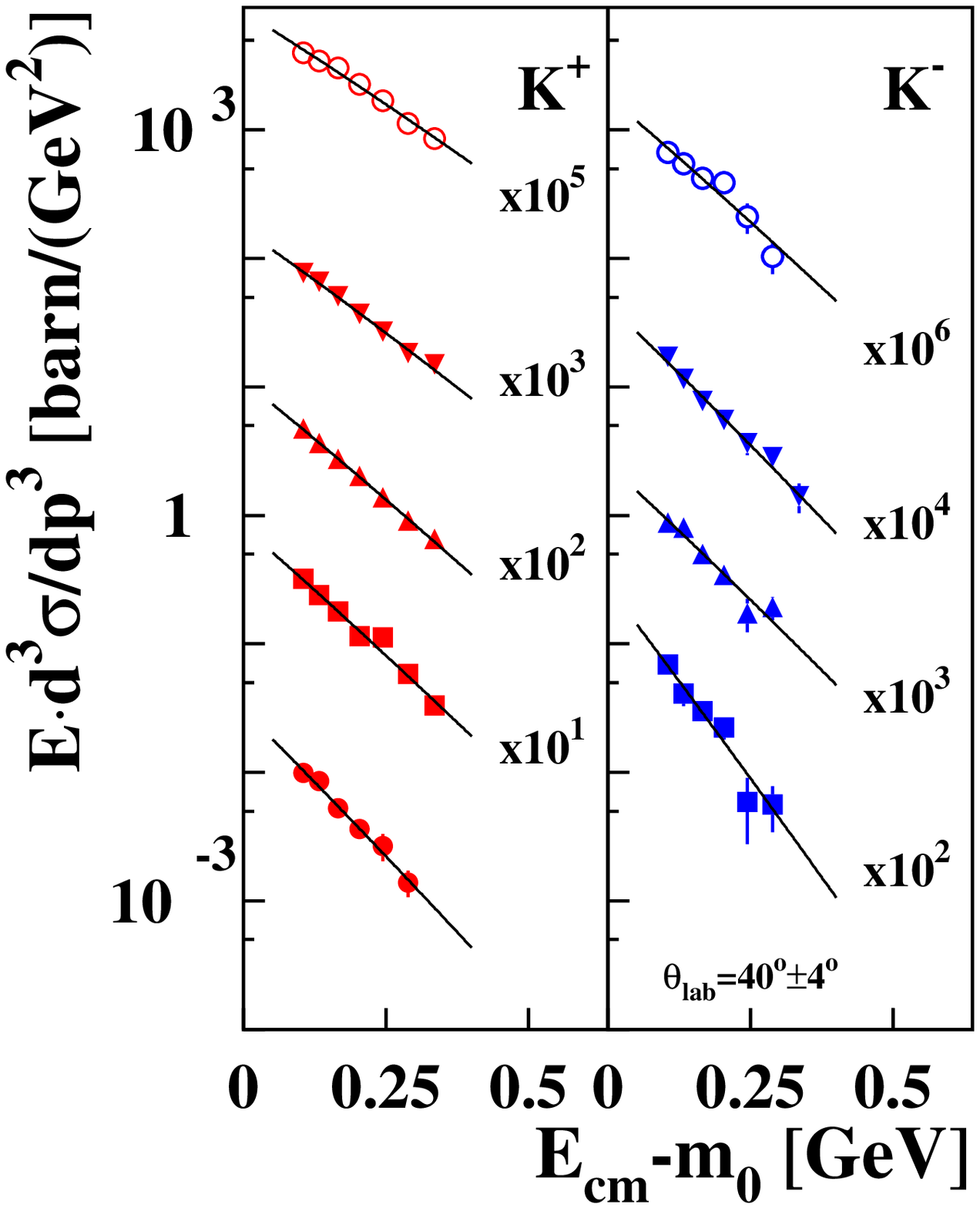,width=6cm,clip=}}
    \mbox{\epsfig{figure=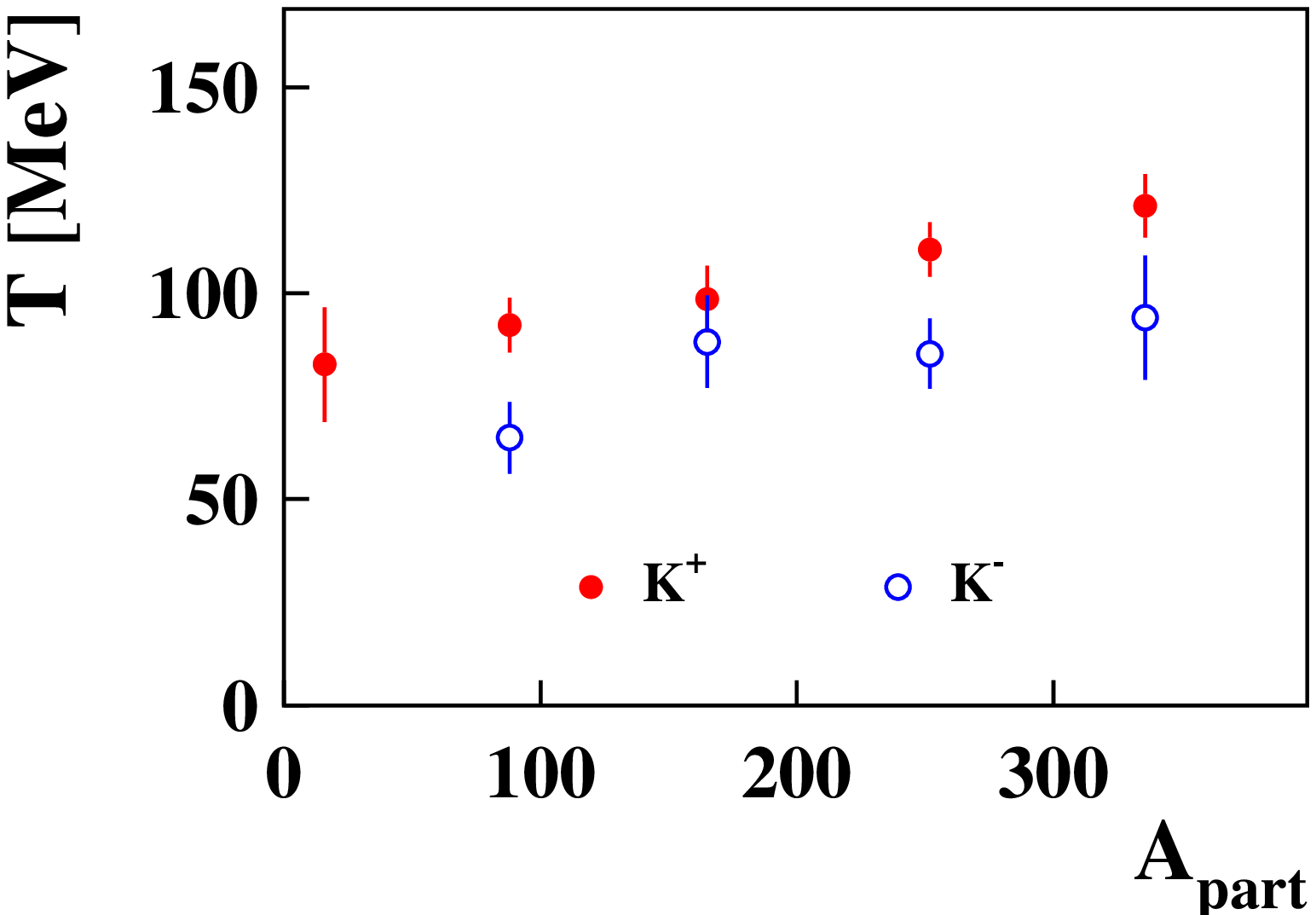,width=6cm,clip=}}
    \caption{{\underline{Left:}}
             Invariant cross sections for K$^+$ and for K$^-$ 
             in \mbox{Au+Au} collisions 
             at $E_{beam}=1.5$~AGeV for the different centrality bins. 
             The open circles depict 
             the most central data, the other bins are shown
             from the top to the bottom 
             of the figure with decreasing centrality.
             {\underline{Right:}} 
             Inverse slope parameters $T$ as a function of the 
             number of participating nucleons $A_{part}$ 
             for K$^+$ (full symbols) and K$^-$ (open symbols).}
    \label{fig_spec_tmbin}
  \end{center} 
\end{figure}

The left hand side of figure \ref{fig_spec_tmbin} shows the production 
cross sections for K$^+$ 
and for K$^-$ mesons measured close to midrapidity as a function of
the kinetic energy in the center-of-momentum system  for the five
centrality bins in Au+Au collisions at $1.5$~AGeV. The uppermost spectra
correspond to the most central reactions, the subsequent bins 
are shown from the top to the bottom 
of the figure with decreasing centrality.  The error bars
represent the statistical uncertainties of the kaon and the
background events. An overall systematic error of 10\% due to
efficiency corrections and beam normalization has to be added. The
solid lines represent the Boltzmann function 
$d^3\sigma/dp^3 = C \cdot E \cdot \exp(-E/T)$ fitted to the data. 
$C$ is a normalization constant and the 
exponential function describes the energy
distribution with $T$ being the inverse slope parameter.

The spectra presented in figure \ref{fig_spec_tmbin} exhibit a distinct
difference between K$^-$ and K$^+$: The slopes of the 
K$^-$ spectra are steeper than those of the K$^+$ spectra. The inverse
slope parameters $T$ as a function of
$A_{part}$ for each centrality bin
are displayed in the right hand side of the figure. 
$T$ increases with increasing centrality and
is found to be significantly lower for K$^-$ than for K$^+$,
even for the most central collisions. The same behaviour
has been observed for \mbox{Ni+Ni} collisions at $1.93$~AGeV.

\begin{figure}[h]
  \begin{center}
  \mbox{\epsfig{figure=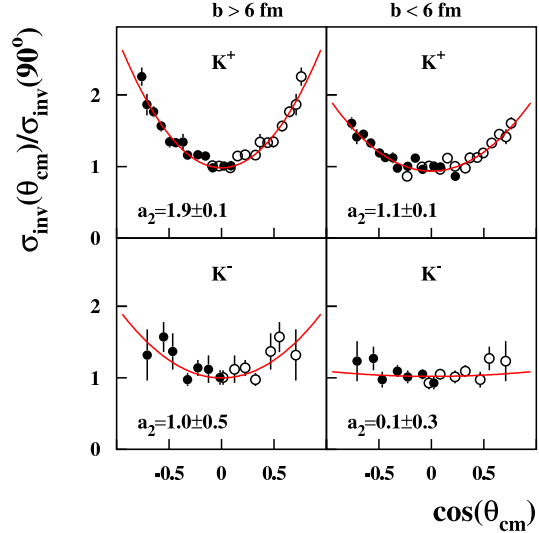,width=7cm,clip=}}
  \caption{Polar angle distributions for K$^+$ (upper panels) 
           and for K$^-$ (lower panels) in \mbox{Au+Au} 
           collisions at $E_{beam}=1.5$~AGeV. The left panels 
           show data for impact 
           parameters $b>6$~fm, the right ones for $b<6$~fm. 
           The fits and the parameter $a_{2}$ 
           are as described in the text.}
  \label{fig_angdis}
\end{center} 
\end{figure}

Another observable showing a distinct difference between 
K$^+$  and K$^-$ is the
polar angle emission pattern.  
Due to
limited statistics we considered only Au+Au collisions at $1.5$~AGeV
grouped into two centrality bins: near-central (impact parameter $b<$6 fm)
and non-central collisions ($b>$6 fm) \cite{foerster}. 
Figure \ref{fig_angdis} displays
the polar angle distribution  for K$^+$ (upper panels) and for K$^-$ (lower
panels), both for near-central (right) and for non-central collisions
(left).

The solid lines represent the function $1 +
a_2 \cdot \cos^2(\theta_{cm})$ which is fitted to the experimental
distributions with the values of $a_2$ given in the figure. In
near-central collisions the K$^-$ mesons  exhibit an isotropic
emission pattern whereas the emission of the K$^+$ mesons is
forward-backward peaked. The angular distributions observed for
K$^+$ and for K$^-$ in \mbox{Ni+Ni} collisions at $1.93$~AGeV are
similar to the ones presented in  figure~\ref{fig_angdis} \cite{menzel}.

\section{The azimuthal distribution of the 
          K$^{\boldsymbol{+}}$- and of the K$^{\boldsymbol{-}}$-emission}
         \label{azimuthal}

\begin{figure}
\begin{center}
  \mbox{\epsfig{file=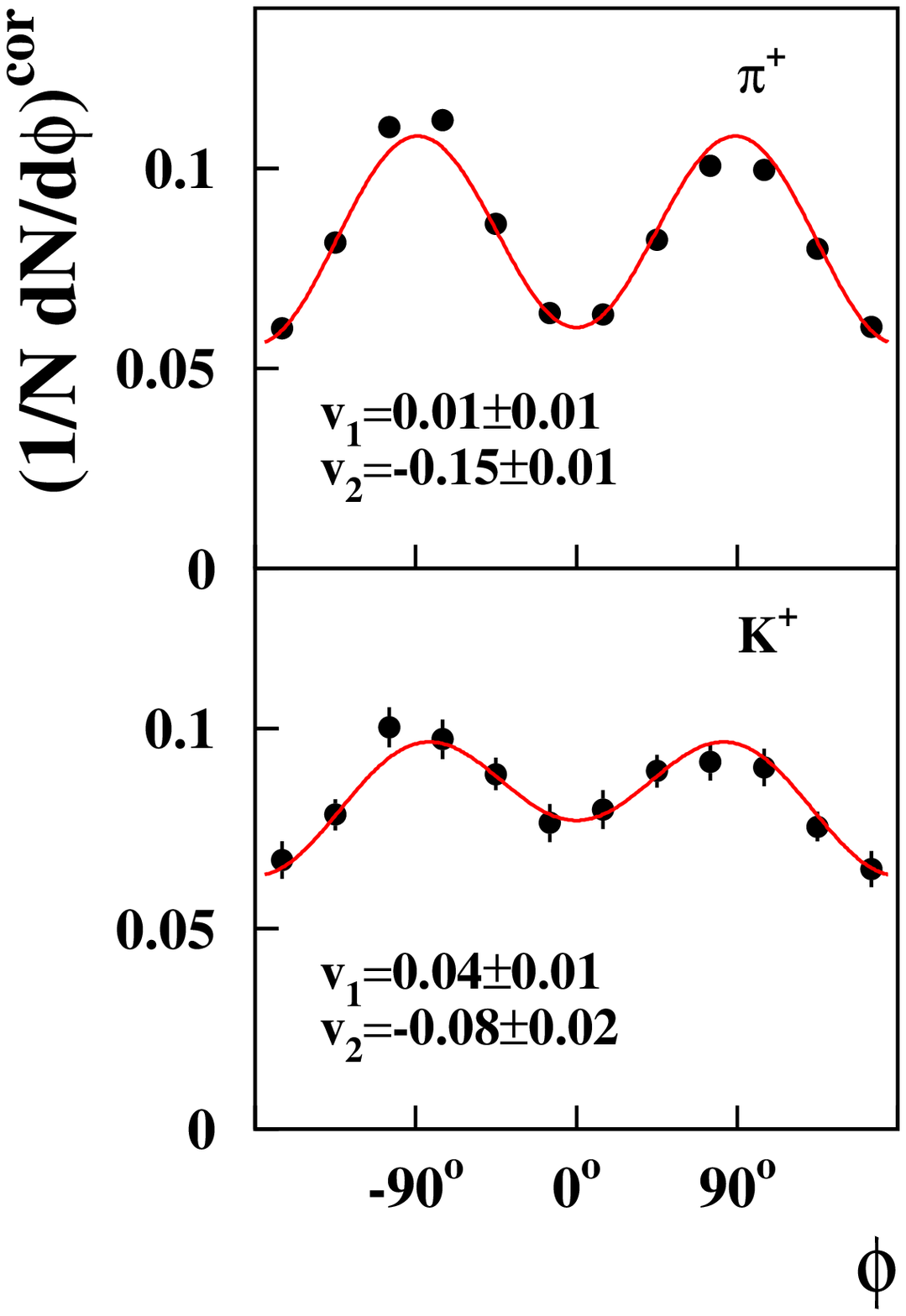,width=6cm}}
  \mbox{\epsfig{file=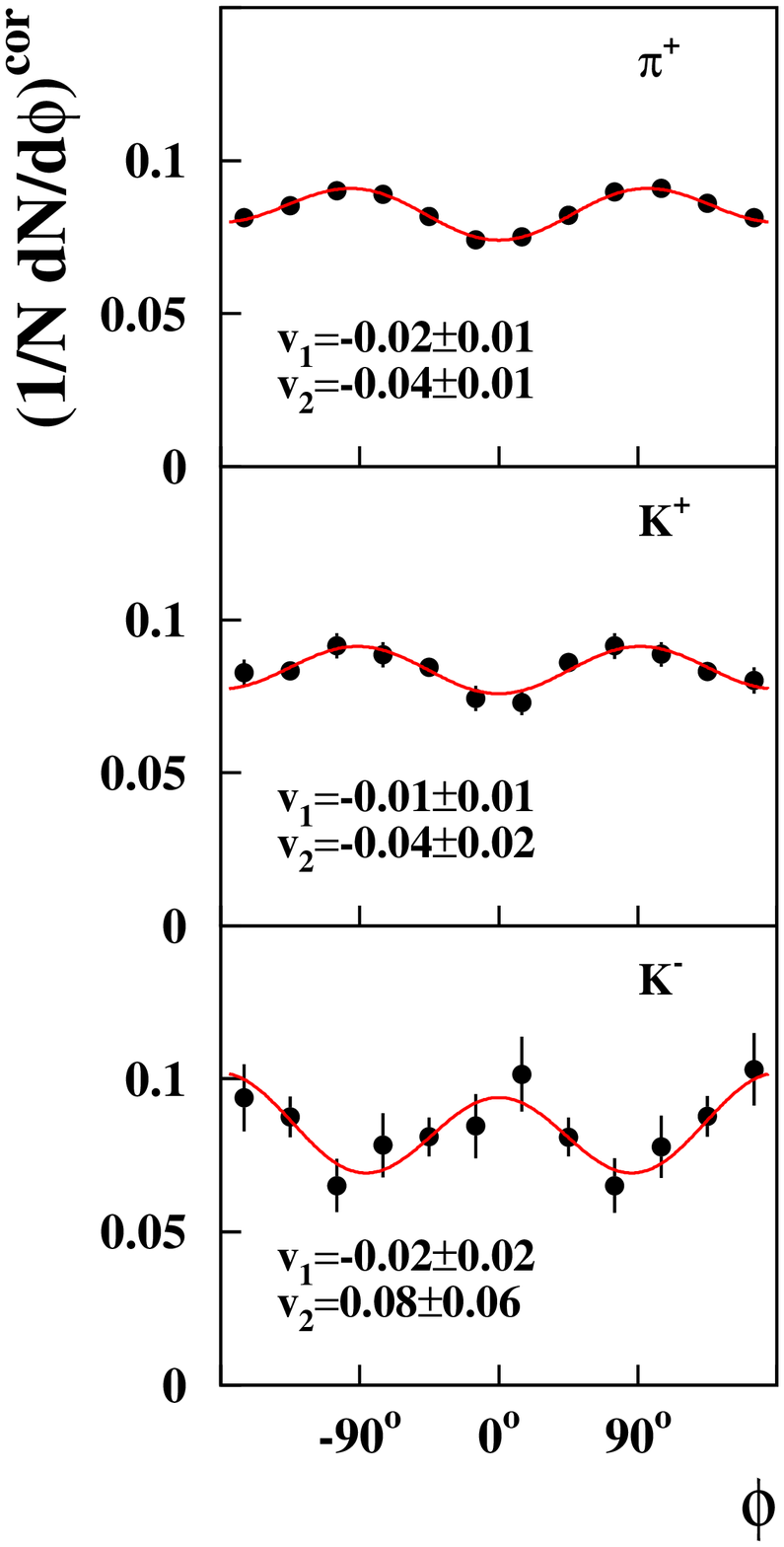,width=6cm}}
  \caption{{\underline{Left:}}
           Azimuthal distributions of $\pi^+$ and of K$^+$ mesons for
           semi-central \mbox{Au+Au} collisions at $1.5$~AGeV. 
           The data are corrected for the resolution of the 
           reaction plane and refer to
           impact parameters of 5.9 fm $< b <$ 10.2 fm, 
           rapidities of 0.3 $<y/y_{beam} <$ 0.7 
           and momenta of  0.2 GeV $< p_t <$ 0.8 GeV.
           The lines are fits as described in the text. 
           {\underline{Right:}}
           Azimuthal distributions of $\pi^+$, of K$^+$ and of 
           K$^-$ mesons for
           semi-central \mbox{Ni+Ni} collisions at $1.93$~AGeV. The data 
           refer to impact parameters of 3.8 fm$ < b <$ 6.5 fm 
           and the same range in rapidity and $p_t$ as the 
           \mbox{Au+Au} data.}
  \label{fig_azimuthal}
\end{center} 
\end{figure}

The left hand side of figure \ref{fig_azimuthal} 
shows the azimuthal distributions of $\pi^+$ 
and K$^+$  mesons for semi-central \mbox{Au+Au} collisions at $1.5$~AGeV,
the right hand side the distributions for $\pi^+$, K$^+$
and K$^-$ in \mbox{Ni+Ni} collisions at $1.93$~AGeV \cite{uhlig}. 
The distributions are corrected for the angular resolution of the
reaction plane determination \cite{brill}.
The data are fitted using the first two components
of a Fourier series
$dN/d\Phi \sim 2\, v_1 \cos(\phi) \, + \, 2\, v_2 \cos (2\phi)$
resulting in values for $v_1$ and $v_2$ as given in the figures
together with the statistical errors.
The determination of the coefficient $v_1$ is subject to an additional
systematic error of 0.04.

In \mbox{Au+Au} both, $\pi^+$ and K$^+$ mesons, exhibit a pronounced
enhancement  at $\phi = \pm 90^o$, i.e.~perpendicular to the
reaction plane. For $\pi^+$ mesons this effect can be interpreted
as rescattering and absorption within the spectator fragments. 

The study of \mbox{Ni+Ni} collisions was performed at a higher
incident energy of $1.93$~AGeV. The resulting higher production cross
section for K$^-$ mesons provides an opportunity to study both charged kaon
species. The data are shown on the right hand side of figure 
\ref{fig_azimuthal} along with $\pi^+$
mesons for semi-central collisions. Both, $\pi^+$ and K$^+$
mesons follow, the same trend already observed in \mbox{Au+Au} collisions.
The values for $v_2$ are
smaller than in \mbox{Au+Au} as one might expect for the smaller system.
In contrast to the $\pi^+$ and to the K$^+$ mesons the K$^-$ mesons 
show an in-plane enhancement.

This ``positive'' (in-plane) elliptic flow of particles is
observed for the first time in heavy-ion collisions at SIS
energies. In contrast to this observation one would expect a
preferential out-of-plane emission (negative elliptic flow) of
K$^-$ mesons due to their large absorption cross section in spectator
matter.

A depletion of the expected out-of-plane emission pattern of K$^-$ mesons
might be due to the fact that they are produced via
strangeness-exchange reactions. This causes a delay in the
freeze-out of the K$^-$ mesons~\cite{foerster,har03,oeschler} and, hence, a
reduced shadowing effect by the spectator fragments which have
moved further away. The observed in-plane emission of K$^-$ mesons,
however, cannot be easily explained by this scenario.

\begin{figure}
\begin{center}
  \mbox{\epsfig{file=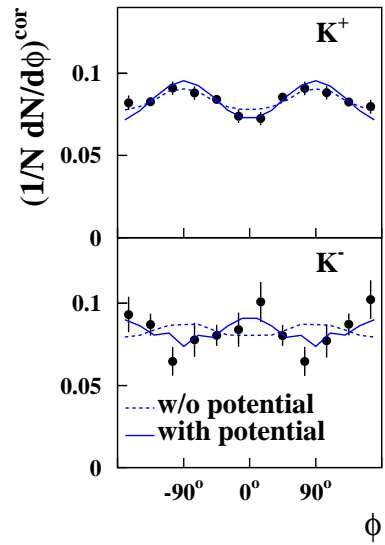,width=6cm}}
  \caption{Comparison of the data from \mbox{Ni+Ni} reactions at $1.93$~AGeV 
           (see fig. \ref{fig_azimuthal}) with IQMD
            model calculations \cite{har_azi}.} 
  \label{fig_iqmd}
\end{center} 
\end{figure}

In order to quantitatively explain the K$^+$ and K$^-$ meson azimuthal
distributions in figure \ref{fig_iqmd} we compare our data
to recent results of the IQMD
model \cite{har_azi}. This transport calculation takes into account
both the space-time evolution of the reaction system and the
in-medium properties of the strange mesons.  The dashed and solid
lines represent results of calculations without and with in-medium
potentials, respectively.  In the case of the K$^+$ mesons 
the effect of the repulsive
K$^+$N potential is small in this model. 
A large fraction of the observed out-of-plane enhancement,
in contrast to other models \cite{wang,li,mishra},
is caused by the scattering of K$^+$ mesons.
The transport code (HSD) \cite{mishra} in contrast
predicts a dominant influence of the potential on the 
emission pattern of the K$^+$ mesons. In the system Au+Au at 1~AGeV where
both size and life time of the fireball are larger than
in the Ni+Ni case, the effect of
the repulsive K$^+$N potential was found to be very important
\cite{shin,wang,li}.

The lower part of figure \ref{fig_iqmd}
shows a comparison of the K$^-$ data to calculations 
without (dashed) and with
(solid) K$^-$N potential.
When neglecting the K$^-$N
potential, the calculation predicts a weak out-of--plane elliptic flow
caused by shadowing. This effect is rather small because of the
late emission of K$^-$ mesons. When taking into account the
attractive in-medium K$^-$N potential the model is able to
describe the experimental in-plane elliptic flow pattern much 
better.
Model calculations with the HSD
code \cite{mishra} predict a flat azimuthal distribution both with
and without a K$^-$N potential and, hence cannot  explain the
observed in-plane flow of K$^-$ mesons.


\section*{References}
\begin{thebibliography}{99}

\bibitem{aich} J. Aichelin and C. M. Ko, Phys. Rev. Lett. {\bf 55} (1985) 2661.
\bibitem{brown} G. Q. Li, C. H. Lee and G. E. Brown, 
                Phys. Rev. Lett. {\bf 79} (1997) 5214.
\bibitem{schaffner} J. Schaffner-Bielich, J. Bondorf, I. N. Mishustin,
                    Nucl. Phys. {\bf A 625} (1997) 325.
\bibitem{senger} P. Senger et al. (KaoS), 
                 Nucl. Instr. Meth. {\bf A 327} (1993) 393.
\bibitem{misko} D. Mi\'skowiec et al. (KaoS), 
                Phys. Rev. Lett. {\bf 72} (1994) 3650.
\bibitem{ahner} W. Ahner et al. (KaoS), Phys. Lett. {\bf B 393} (1997) 31.
\bibitem{barth} R. Barth et al. (KaoS), Phys. Rev. Lett. {\bf 78} (1997) 4007.
\bibitem{shin} Y. Shin et al. (KaoS), Phys. Rev. Lett. {\bf 81} (1998) 1576.
\bibitem{laue} F. Laue, C. Sturm et al. (KaoS), 
               Phys. Rev. Lett. {\bf 82} (1999) 1640.
\bibitem{menzel} M. Menzel et al. (KaoS), Phys. Lett. {\bf B495} (2000) 26.
\bibitem{sturm} C. Sturm et al. (KaoS), Phys. Rev. Lett. {\bf 86} (2001) 39.
\bibitem{foerster} A. F{\"o}rster, F. Uhlig et al. (KaoS), 
               Phys. Rev. Lett. {\bf 91} (2003) 152301. 
\bibitem{uhlig}  F. Uhlig, A. F{\"o}rster et al. (KaoS), 
                submitted to Phys. Rev. Lett., nucl-ex/0411021.
\bibitem{li_ko} G.Q. Li and C.M. Ko, Phys. Rev. {\bf C 54} (1996) R2159.
\bibitem{aichelin} J. Aichelin, Phys. Rep. {\bf 202} (1991) 233.
\bibitem{fuchs} C. Fuchs et al., Phys. Rev. Lett. {\bf 86} (2001) 1974.
\bibitem{har02} C. Hartnack and J. Aichelin, J. Phys. G {\bf 28} (2002) 1649.
\bibitem{massaction} J. Cleymans, A. F{\"orster}, H. Oeschler, K. Redlich,
                     F. Uhlig , Phys. Lett. {\bf B 603} (2004) 146.
\bibitem{brill} D. Brill et al. (KaoS), Z. Phys. {\bf A 355} (1996) 61.
\bibitem{har03} C. Hartnack, H. Oeschler, J. Aichelin, 
                 Phys. Rev. Lett. {\bf 90} (2003) 102302; 
                 Phys. Rev. Lett. {\bf 94} (2004) 149903.
\bibitem{oeschler} H. Oeschler, J. Phys. G {\bf 27} (2001) 257.
\bibitem{har_azi} C. Hartnack et al., Eur. Phys. J. {\bf A 1} (1998) 151 
                   and to be published.
\bibitem{wang} Z.S. Wang, C.Fuchs, A. Faessler, T. Gross-Boelting, 
               Eur. Phys. J. {\bf A 5} (1999) 275.
\bibitem{li} G. Q. Li et al., Phys. Lett. {\bf B 381} (1996) 17.
\bibitem{mishra} A. Mishra, E. Bratkovskaya, J. Schaffner-Bielich,
                 S. Schramm, H. St{\"o}cker, 
                 Phys. Rev. {\bf C 70} (2004) 044904.

\end{thebibliography}
\end{document}